\documentstyle[aps,epsfig,preprint]{revtex}
\begin{document}
\tightenlines
\draft
\preprint{}
\title{Numerical solution of the color superconductivity gap in a weak coupling constant}
\date{today}
\author{I. Zakout}
\address{Department of Physics, Stanford University, Stanford, CA 94305-4060}
\author{H.R. Jaqaman} 
\address{Department of Physics, Bethlehem University, P.O. Box 9, Bethlehem, PA}   
\author{W. Greiner}
\address{Institut f\"ur Theoretische Physik, Robert Mayer Str. 8-10,
D-60054, Frankfurt am Main, Germany}

\maketitle
\begin{abstract}
We present the numerical solution of the full gap equation in a weak coupling constant $g$. 
It is found that the standard approximations to derive the gap equation 
to the leading order of coupling constant are essential for a secure numerical evaluation 
of the logarithmic singularity with a small coupling constant.
The approximate integral gap equation with a very small $g$ should be inverted 
to a soft integral equation to smooth the logarithmic singularity near the Fermi surface.
The full gap equation is solved for a rather large coupling constant $g\ge 2.0$.
The approximate and soft integral gap equations are solved for small $g$ values.
When their solutions are extrapolated to larger $g$ values, they coincide the full gap equation solution near the Fermi surface. 
Furthermore, the analytical solution matches the numerical one up to the order one $O(1)$. 
Our results confirm the previous estimates that the gap energy is of the order tens to 100 MeV for the chemical potential 
$\mu\le 1000$ MeV. 
They also support the validity of leading approximations applied to the full gap equation to derive the soft integral gap equation 
and its analytical solution near the Fermi surface.  
\end{abstract}
\pacs{ }

\mediumtext
\section{Introduction}
The deconfined quark matter is expected at baryon densities much larger than the normal nuclear matter density. 
Any attractive quark-quark interaction causes instability of the Fermi surface by formation of Cooper pairs and leads to the color 
superconductivity phase\cite{Barrois1,Frautschi1,Bailin1} in a very dense quark matter at low temperatures. 
However, the quark matter at relatively low densities may have properties qualitatively different from those for BCS-type 
superconductor. It is argued that the nuclear matter at low density might be continuously connected to the quark matter at high 
density without any phase transition. The calculation of superconducting gap is important to study many properties of 
hadronic matter at high densities\cite{Alford01,Rajagopal00}.

In QCD, the single-gluon exchange between quarks of different color generates an attractive interaction in the color antitriplet 
channel. The scattering through a single-gluon exchange strongly correlates the direction of the in- and outgoing quarks. 
However, there is a logarithmic divergence for the forward angle scattering.
This implies that the gap equation is not an exponential in $1/g^2$\cite{Bailin1}, as in BCS like theories, but only in 
$1/g$\cite{Son1} where $g$ is the coupling constant.
It is found for the nonzero quark density at low temperature, the free energy is an expansion in $g^2\ln(1/g)$, and appears well 
behaved even for much larger coupling constant $g\le 4$\cite{Schafer1}.

The earlier estimations of the color superconductivity gap were found rather small $\Delta\approx \mbox{(few) MeV}$ for quark 
chemical potentials that are of physical interest $\mu\le 1000 MeV$\cite{Barrois1,Frautschi1,Bailin1}. 
Recently, it is shown that the gap is large and of order 100 MeV for $\mu < 1000$ MeV \cite{Schafer1,Alford97a,Rapp97a}. 
These large estimations may lead to observable effects in the core of neutron stars and the physics of heavy ion 
collisions\cite{Alford01,Rajagopal00,AlfordCFL}.
The reliable estimation of gap equation plays an essential role in the phenomenological calculations involve the color 
superconductivity physics.

The gap equation has been studied numerically for a quantitatively estimation for a wide range of the chemical potential in the 
time-like energy space\cite{Schafer1} as well as momentum space\cite{Abuki,Fugleberg1,Fugleberg2}. 
In these studies, the gluon propagator has been specified by Thomas-Fermi screening and Landau damping defined by the hard dense 
loop approximation (HDL).  
Furthermore, the resultant equation has been simplified by approximating the angular integration near the Fermi surface\cite{Abuki}.
Nonetheless, the gap equation is highly singular near the Fermi surface due to the correlation of in- and outgoing quarks direction. 
Therefore, any approximation may affect significantly the gap value in particular for a moderate quark chemical potential.
In the present work, we shall adopt the spectral representation of the gluon propagator\cite{Pisarski1,Pisarski2} 
and solve the gap equation exactly numerically without any further approximation.  
The spectral representation is more rigorous than previous approximation based on screening gluon propagator.

Recently, Pisarski and Rischke\cite{Pisarski1,Pisarski2,Rischke1,Rischke2} have derived the full gap equation in the momentum space 
for two massless quark flavors with a total spin $J=0$ at an arbitrary temperature $T$.
They adopted the gluon propagator in the spectral representation derived from HDL approximation. They derived the approximate gap 
equation to the leading order of coupling constant $g$. It is given by Eq.(72) in Ref.\cite{Pisarski1}. 
This equation has a logarithm dependence on the quark energy under the integral.
Furthermore, they have approximated the logarithmic term under the integral to derive a soft integral equation that is given by 
Eq.(80) in Ref.\cite{Pisarski1}.
Hereafter, the approximated integral gap equation is called Eq. (I) while the soft integral gap equation is called Eq. (II).
Finally, they have inverted the soft integral equation to an ordinary differential equation to derive the analytical solution.

In the present work, the full gap equation is solved numerically for a rather large coupling constant $g=2-4$ which corresponds 
to the chemical potential $\mu < 1000$ MeV. 
We confirm the previous estimates that for the chemical potential of order 1 GeV, the gap amplitude can be of order tens to 100 MeV.
We also study the numerical solution for a small coupling constant $g<2.5$ by solving Eq. (I) as well as Eq. (II).
We compare the numerical solutions with the analytical one to the leading order logarithmic accuracy.
This paper is organized as follows.
In Sec. II, we present the full gap equation and its approximated versions.
The results and the conclusions are presented in Sec. III.
%%%%%%%%%%%%%%%%%%%%%%%%%%%%%%%%%%%%%%%%%%%%%%%%%%%%%%%%%%%%%%%%%%%%%%%%%%%%%%

\section{The gap equation} 
The full gap equation for the color superconductivity in the weak coupling with total spin zero and two massless flavors in a dense 
QCD reads \cite{Pisarski1}
\begin{eqnarray}
\phi(k)=
\frac{2}{3} g^{2} \int \frac{d^{3}q}{(2\pi)^{3}}
\left[
T\sum_{q_0}\Delta_l(K-Q)\Xi(Q){\cal P}_l(\hat{\bf k}\cdot\hat{\bf q})
+T\sum_{q_0}\Delta_t(K-Q)\Xi(Q){\cal P}_t(\hat{\bf k}\cdot\hat{\bf q})
\right]
\end{eqnarray}
where
\begin{eqnarray}
{\cal P}_l(\hat{\bf k}\cdot\hat{\bf q})=
\left(\frac{1+\hat{\bf k}\cdot\hat{\bf q}}{2}\right)
\end{eqnarray}
and
\begin{eqnarray}
{\cal P}_t(\hat{\bf k}\cdot\hat{\bf q})=
\left(
-\frac{3-\hat{\bf k}\cdot\hat{\bf q}}{2}
+\frac{1+\hat{\bf k}\cdot\hat{\bf q}}{2}
\frac{(k-q)^2}{({\bf k}-{\bf q})^2}
\right).
\end{eqnarray}
The Matsubara sums over $q_0$ in the spectral representation for the quantity $\Xi(Q)$ are
computed as
\begin{eqnarray}
&T&\sum_{q_0}\Delta_l(K-Q)\Xi(Q)=
-\frac{\phi(\epsilon_q,{\bf q})}{2\epsilon_q}
\left\{-\frac{1}{2}\tanh\left(\frac{\epsilon_q}{2T}\right)\frac{2}{p^2}
\right. \nonumber\\
&+&
\int_{0}^{\infty}d\omega \rho_l(\omega,{\bf p})
\frac{1}{2}\tanh(\frac{\epsilon_q}{2T})
\left[
\frac{1}{\epsilon_k+\epsilon_q+\omega}
-\frac{1}{\epsilon_k-\epsilon_q-\omega}
-\frac{1}{\epsilon_k+\epsilon_q-\omega}
+\frac{1}{\epsilon_k-\epsilon_q+\omega}
\right]
\nonumber\\
&+&
\left.
\int_{0}^{\infty}d\omega \rho_l(\omega,{\bf p})
\frac{1}{2}\coth(\frac{\omega}{2T})
\left[
\frac{1}{\epsilon_k+\epsilon_q+\omega}
-\frac{1}{\epsilon_k-\epsilon_q-\omega}
+\frac{1}{\epsilon_k+\epsilon_q-\omega}
-\frac{1}{\epsilon_k-\epsilon_q+\omega}
\right]\right\},
\end{eqnarray}
and
\begin{eqnarray}
&T&\sum_{q_0}\Delta_t(K-Q)\Xi(Q)=
\nonumber\\
&-&\frac{\phi(\epsilon_q,{\bf q})}{2\epsilon_q}
\left\{
\int_{0}^{\infty}d\omega \rho_t(\omega,{\bf p})
\frac{1}{2}\tanh(\frac{\epsilon_q}{2T})
\left[
\frac{1}{\epsilon_k+\epsilon_q+\omega}
-\frac{1}{\epsilon_k-\epsilon_q-\omega}
-\frac{1}{\epsilon_k+\epsilon_q-\omega}
+\frac{1}{\epsilon_k-\epsilon_q+\omega}
\right] \right.
\nonumber\\
&+&
\left.
\int_{0}^{\infty}d\omega \rho_t(\omega,{\bf p})
\frac{1}{2}\coth(\frac{\omega}{2T})
\left[
\frac{1}{\epsilon_k+\epsilon_q+\omega}
-\frac{1}{\epsilon_k-\epsilon_q-\omega}
+\frac{1}{\epsilon_k+\epsilon_q-\omega}
-\frac{1}{\epsilon_k-\epsilon_q+\omega}
\right]\right\},
\end{eqnarray}
where $\epsilon_q=\sqrt{ (q-\mu)^2+|\phi(q)|^2 }$.
The spectral densities, $\rho_{l}$ and $\rho_{t}$, are given by
\begin{eqnarray}
\rho_{l,t}(\omega,{\bf p})=
\rho^{\mbox{pole}}_{l,t}(\omega,{\bf p})
\delta[\omega-\omega_{l,t}({\bf p})]
+\rho^{\mbox{cut}}_{l,t}(\omega,{\bf p})\theta(p-\omega),
\end{eqnarray}
The explicit expressions for the spectral densities
$\rho^{\mbox{pole,cut}}_{l,t}$
are given by Eqns(34b-e) in Ref. \cite{Pisarski1}.
The $\omega_{l,t}({\bf p})$ are the solutions of the dispersion relations for longitudinal and transverse gluons which are given by 
Eqns(37a-b) in Ref.\cite{Pisarski1}.

In HDL approximation there are three scales in cold dense QCD, the chemical potential $\mu$, the gluon mass $m_g\sim g\mu$ and the 
color superconductivity condensate $\phi_0\sim \mu\exp(-1/g)$ and they are naturally ordered $\mu>>m_g>>\phi_0$.
Pisarski and Rischke \cite{Pisarski1} have derived the approximate integral gap equation by using the assumptions of these scales 
from the full gap equation. This approximate integral gap equation is derived in the momentum space
unlike to the Eliashberg equation given by 
Sch\"afer and Wilczak \cite{Schafer1} is derived in the time-like energy space.

The approximate integral equation, Eq.(I), reads
\begin{eqnarray}
\phi(k)\approx\frac{g^{2}}{18\pi^2}\frac{1}{2}\int^{\mu+\delta}_{\mu-\delta}
\frac{dq}{\epsilon_q}\tanh\left(\frac{\epsilon_q}{2T}\right)
\frac{1}{2}\ln\left(\frac{b^2\mu^2}{|\epsilon^2_q-\epsilon^2_k|}\right)
\phi_q,
\end{eqnarray}
where $b=256\pi^{4}\left(\frac{2}{N_fg^2}\right)^{5/2}$. 
This equation can be simplified to a soft integral equation by replacing
the logarithm under the integral by \cite{Son1} 
\begin{eqnarray}
\frac{1}{2}\ln\left(\frac{b^2\mu^2}{|\epsilon^2_q-\epsilon^2_k|}\right)
\approx 
\ln\left(\frac{b\mu}{\epsilon_q}\right)\theta(q-k)+
\ln\left(\frac{b\mu}{\epsilon_k}\right)\theta(k-q).
\end{eqnarray}
Hence, the soft integral gap equation, Eq.(II), becomes
\begin{eqnarray}
\phi(k)\approx\overline{g}^2
\left[
\ln\left(\frac{b\mu}{\epsilon_k}\right)
\int^{k}_{\mu}\frac{dq}{\epsilon_q}
\tanh\left(\frac{\epsilon_q}{2T}\right)\phi_q 
+\int^{\mu+\delta}_{k}
\frac{dq}{\epsilon_q}
\tanh\left(\frac{\epsilon_q}{2T}\right)
\ln\left(\frac{b\mu}{\epsilon_q}\right)
\phi_q
\right]
\end{eqnarray}
where $\overline{g}=\frac{g}{3\sqrt{2}\pi}$.

The soft integral gap equation, Eq.(II), is inverted to an ordinary differential equation which has a simple solution to the 
leading order in $\overline{g}$ and Fermi surface $x^{*}=\pi/2\overline{g}$.
The solution for zero temperature case reads
\begin{eqnarray}
\phi(x)=2b\mu\exp\left(-\pi/(2\overline{g}) \right)
\sin(\overline{g}x) \times {\cal O}(1) 
\end{eqnarray}
where $x=\ln\left(\frac{2b\mu}{k-\mu+\epsilon_k}\right)$.
Furthermore, the analytical solution of the gap equation in the time-like energy space, Eliashberg equation, is given by \cite{Schafer1}
\begin{eqnarray}
\phi_0\approx b\mu\exp(-\frac{\pi}{2\overline{g}}) \times {\cal O}(1)
\end{eqnarray}
where the overall coefficient is correct up to a prefactor of order 
one ``${\cal O}(1)$''.
Since the analytical solution of the gap equation to leading order of coupling constant $g$ is undetermined by a prefactor of order 
one, we assume that the analytical solution at the Fermi surface reads
\begin{eqnarray}
\phi_0=\zeta b\mu\exp(-\frac{\pi}{2\overline{g}})
\end{eqnarray}
where $\zeta$ is constant of order one. 

%%%%%%%%%%%%%%%%%%%%%%%%%%%%%%%%%%%%%%%%%%%%%%%%%%%%%%%%%%%%%%%%
%%%%%%%%%%%%%%%%%%%%%%%%%%%%%%%%%%%%%%%%%%%%%%%%%%%%%%%%%%%%%%%%
\section{Results and conclusions} 
We have studied the numerical solution of the full gap equation and the approximations validity for Eqns. (I) and (II) 
near the Fermi surface.
The full gap equation is solved in momentum space for a rather large coupling constant $g>2.0$. 
The approximate versions of the full gap equation (i.e. Eq.(I) and (2)) are solved for small coupling constant values $g<1$.
Furthermore, we have assumed that Eqns. (I) and (II) are extrapolated to large coupling constant values and then we have presented 
their numerical solutions for coupling constant values $1.0<g<3.0$. 

Fig. 1 displays the relative gap amplitude $\phi(|{\bf k}|)/\phi_0$ versus the scaled Fermi momentum $k/\mu$.
It is shown that the solution of Eq. (II) have a sharp peak near the Fermi surface for small coupling constant values $g\le 1$.
The peak around the Fermi surface for Eq.(II) is much more sharp than that for Eq.(I) with 
a small coupling constant $g\le 1$. 
Hence, the replacement of logarithm under the integral in Eq.(I) by step functions in Eq. (II) secures 
the singularity problem more thoroughly than the Cauchy numerical integration in Eq.(I) for a small coupling constant $g\le 1.0$.
For the large coupling constant, $g>1$, both equations (I) and (II) give similar results. 
As we shall show below, Eq. (I) gives much a smaller amplitude than that for Eq. (II) with a small coupling constant $g\le 1$.
The reason for that is simple. 
The gap equation appears very singular near the Fermi surface for a small coupling constant. 
Hence the approximation in Eq. (II) evaluates the gap amplitude value near the Fermi surface more thoroughly than numerical Cauchy 
integration.
Furthermore, the gap amplitude in Eq. (II) approaches the same behavior for 
that in Eq. (I) with large coupling constants $g>1$. 
Therefore, the numerical Cauchy integration in Eq. (I) becomes unreliable for a very small coupling constant $g<1$. 

Fig. 2 displays the gap amplitude $\phi/\mu$ for the full gap equation versus the relative momentum $k/\mu$ with rather large 
coupling constant values $g>2.0$ that correspond chemical potentials of the order $1$ GeV. 
It is shown that the gap amplitude has a peak near Fermi surface and the function becomes more smooth as the coupling constant 
$g$ increases.

Fig. 3 displays the gap amplitude versus the logarithmic parameter $x$ for the gap Eqs. (I) and (II) as well as the analytical 
solution normalized to the amplitude of Eq. (II). 
The maximum value represents the gap amplitude near Fermi surface. 
Hence the values of $\phi(x_0)/\mu$ and $x_0$ are important near Fermi surface $k=\mu$.
It is shown that for small coupling constant values $g<1$, the analytical solution and numerical solution of Eq. (II) have almost 
the same behavior while the numerical solution of Eq. (I) has a significant deviation.
As the coupling constant $g$ increases, the numerical solution of (I) and (II) approaches each other and almost coincides 
for $g\ge 1.5$. 
However, the analytical solution deviates from the numerical one (II) significantly as the coupling constant exceeds $g>1.5$.
This means that Eq. (I) and Eq. (II) becomes almost the same for a large $g$ and the standard approximation 
which are applied to Eq. (II) to derive the analytical solution fails at large coupling constant.  

In Fig. 4, the gap amplitude $\phi(x)/\mu$ versus the logarithmic parameter
$x$ with coupling constant $g=2.5$ is displayed for the full gap equation as well as for  Eqns. (I) and (II).
It is shown that the amplitude of the full gap equation approaches the values of Eqns. (I) and (2) more closely near the 
Fermi surface 
(i.e. at the hill of the curves) while it deviates significantly for momentum far away from the Fermi surface. 
Eqns. (I) and (II) solutions have almost similar behavior for the most values of $x$. 
It is interesting to note here that the validity of Eqns. (I) and (II) is restricted only near the Fermi surface and any solution 
far away from the Fermi surface isn't guaranteed. 
Hence, Eqns. (I) and (II) give asymptotically accurate solutions near the Fermi surface for large coupling constant values $g\ge 2$.
However, Eq. (I) solution seems accurate for a moderate coupling constant $g$ since the singularity near the Fermi surface becomes 
less severe. 
In the other hand, the solution of Eq. (II) seems more accurate than that for Eq. (I) for small coupling constant since it seems to 
account the singularity near Fermi surface more precisely.

Fig. 5 displays the gap amplitude $\phi_0/\mu$ at the Fermi surface versus the coupling constant $g$ for the full gap equation and 
its approximation to Eqns. (I) and (II) as well as the analytical solution with prefactors $\zeta=0.5$ and $1.0$.  
We have shown the solutions for Eqns. (I) and (II) with $g\ge 0.7$ and the solution of the full gap equation with $g>2.0$.
The sharp peak near the Fermi surface for Eq. (II) is larger than that for Eq. (I) for a coupling constant $g<1$ since 
Eq. (II) is expected to secure the singularity near the Fermi surface better than Eq. (I) as mentioned above. 
However, the solutions (I) and (II) become close to each other for a wide range of coupling constant values $1<g<3$.
The solutions of Eqns. (I) and (II) match the solution of full gap equation for $2<g<3.0$. 
They deviate from the full gap equation for $g>3.0$. 
The full gap equation solution, for $g>3.0$, starts to increase linear logarithmic with respect to $g$.
We also display the analytical solution with prefactor coefficients $\zeta=0.5$ and $2.0$.
The analytical solution with $\zeta=2.0$ seems to fit the solution of Eq. (I) for small $g<1$. 
When the coupling constant $g$ increases, the analytical solution with a prefactor coefficient $zeta=1/2$ seems 
to fit the solutions of Eqns. (I) and (II) as well as the full gap equation for a range $1<g<3.0$. 
However, the analytical solution fails to fit the full gap one 
with a specific prefactor coefficient for the large coupling constant range $g>3.0$. 
Therefore, the analytical solution approximation breaks down for a rather large coupling constant $g>3.0$. 

Fig. 6 displays the amplitude of full gap equation $\phi_0$ at the Fermi surface versus the chemical potential $\mu$ 
in the range 550 MeV $<\mu<$ 800 MeV.
We have used the running coupling constant defined by $g^2=(8\pi^2/\beta_0)\times 1/\ln(\mu^{2}/\Lambda^{2}_{QCD})$
where $\beta_0=(11 N_c - 2 N_f)/3$. 
The QCD scale is taken as $\Lambda_{QCD}=400$ MeV.
The value of the gap amplitude is found of the order of tens to 100 MeV with quark chemical potential that may exist in neutron stars. 
This large value of the gap amplitude near the Fermi surface confirms the earlier estimates\cite{Schafer1}.

%%%%%%%%%%%%%%%%%%%%%%%%%%%%%%%%%%%%%%%%%%%%%%%%%%%%%%%%%%%%%%%%%%%%%%%%%%%%%%
%%%%%%%%%%%%%%%%%%%%%%%%%%%%%%%%%%%%%%%%%%%%%%%%%%%%%%%%%%%%%%%%%%%%%%%%%%%%%%
In summary, we have studied the numerically solution of the full gap equation and its approximations. The numerical results are 
compared to the analytical one with the leading order $g$ approximation.
It is found that the gap equation has a sharp peak near the Fermi surface for a small coupling constant $g<1$.
The singularity at the Fermi surface is hard to secure numerically in the full gap equation with small $g$. 
The approximation to a leading coupling constant $g$ near the Fermi surface is essential to smear and evaluate this singularity 
in Eqns. (I) and (II).
When Eqns. (I) and (II) are extrapolated to large coupling constant values $g>2$, they match the results of the full gap equation 
near the Fermi surface. 
It is shown that the analytical solution coincides the solution of Eq. (II) for small $g$.
However, this analytical solution approaches those of Eq. (I) and full gap equation with a rather large coupling constant $g\le 1.5$ 
only near Fermi surface.
This justifies the validity of approximation hold in Eqns. (I) and (II) near Fermi surface.

\acknowledgments
We would like to thank  D. H. Rischke and M. G. Alford for the discussions.
Financial support by the Deutsche Forschungsgemeinschaft through the
grant GR 243/51-2 is gratefully acknowledged. 
One of us (I.Z.) thanks Fulbright Foundation for the financial support.

%%%%%%%%%%%%%%%%%%%%%%%%%%%%%%%%%%%%%%%%%%%%%%%%%%%%%%%%%%%%%%%%%%%%%%%%%%%%%%%%%%%%%%%%%%%
%%%%%%%%%%%%%%%%%%%%%%%%%%%%%%%%%%%%%%%%%%%%%%%%%%%%%%%%%%%%%%%%%%%%%%%%%%%%%%%%%%%%%%%%%%%

%%%%%%%%%%%%%%%%%%%%%%%%%%%%%%%%%%%%%%%%%%%%%%%%%%%%%%%%%%%%%%%%%%%%%%
%%%%%%%%%%%%%%%%%%%%%%%%%%%%%%%%%%%%%%%%%%%%%%%%%%%%%%%%%%%%%%%%%%%%%%
% fig1
%
\begin{figure}
\centerline{\mbox{\epsfig{file=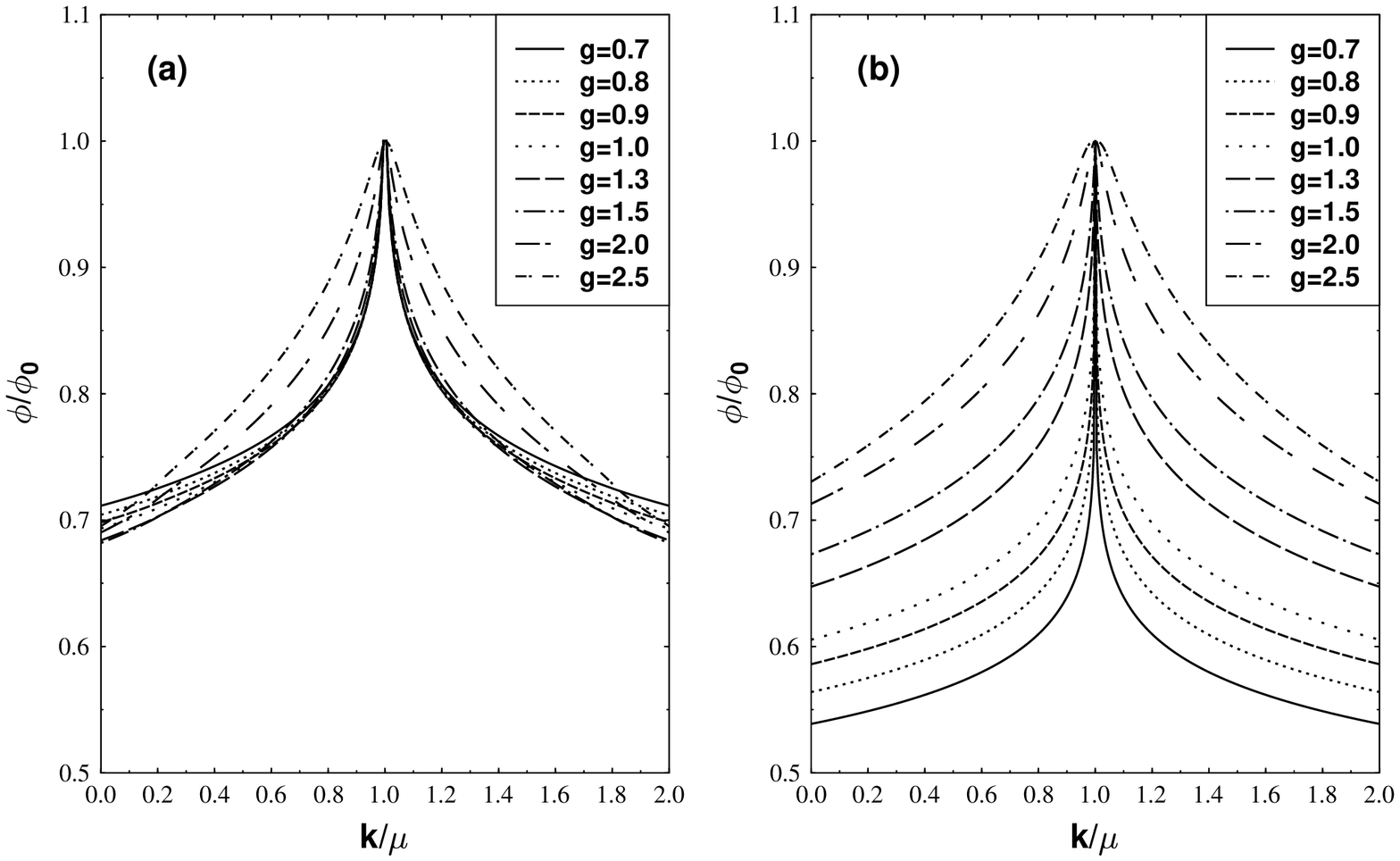,angle=-90,width=1.0\linewidth}}}
\vspace{1truein}
\caption{The color superconductivity gap amplitude $\phi/\phi_0$ versus $k/\mu$ with various coupling constant $g$.
There is a peak at the Fermi surface momentum $k/\mu=1$.
(a) Eq. (I). (b) Eq. (II).}
\label{fig1}
\end{figure}
%%%%%%%%%%%%%%%%%%%%%%%%%%%%%%%%%%%%%%%%%%%%%%%%%%%%%%%%%%%%%%
%%%%%%%%%%%%%%%%%%%%%%%%%%%%%%%%%%%%%%%%%%%%%%%%%%%%%%%%%%%%%
% fig2
%
\begin{figure}
\centerline{\mbox{\epsfig{file=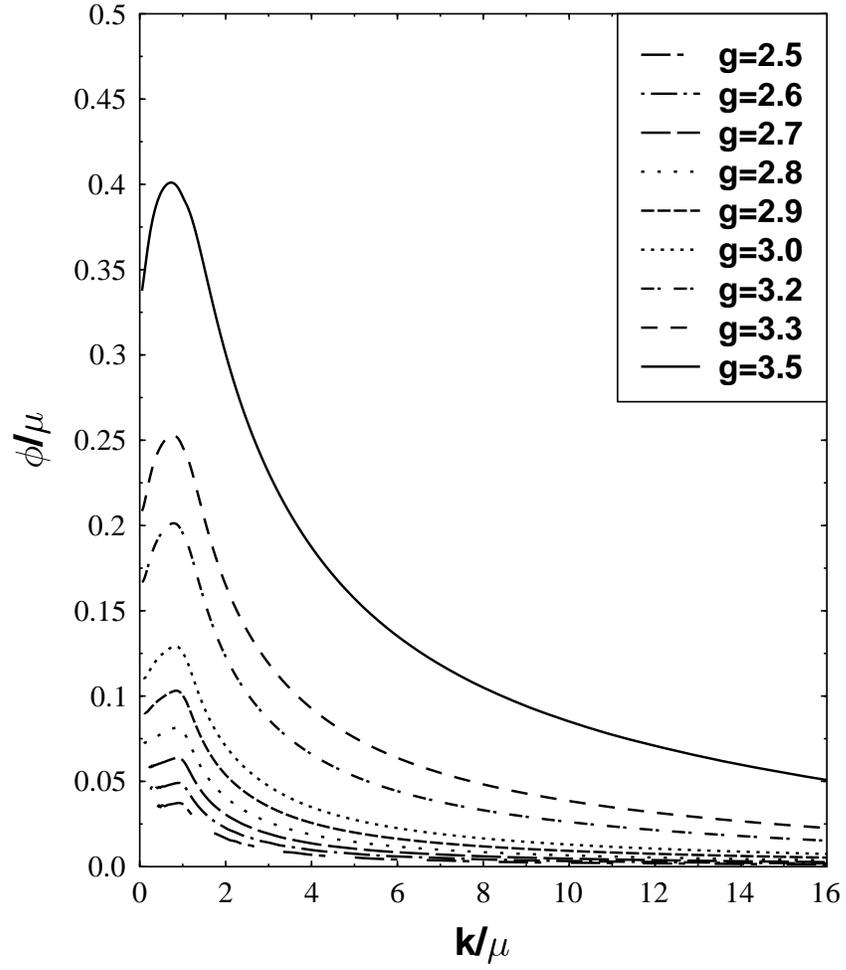,angle=0,width=0.8\linewidth}}}
\vspace{1truein}
\caption{The color superconductivity gap amplitude $\phi_0/\mu$ versus $k/\mu$ for the full gap equation with various coupling 
constant $g$.}
\label{fig2}
\end{figure}
%%%%%%%%%%%%%%%%%%%%%%%%%%%%%%%%%%%%%%%%%%%%%%%%%%%%%%%%%%%%%%
%%%%%%%%%%%%%%%%%%%%%%%%%%%%%%%%%%%%%%%%%%%%%%%%%%%%%%%%%%%%%
% fig3
%
\begin{figure}
\centerline{\mbox{\epsfig{file=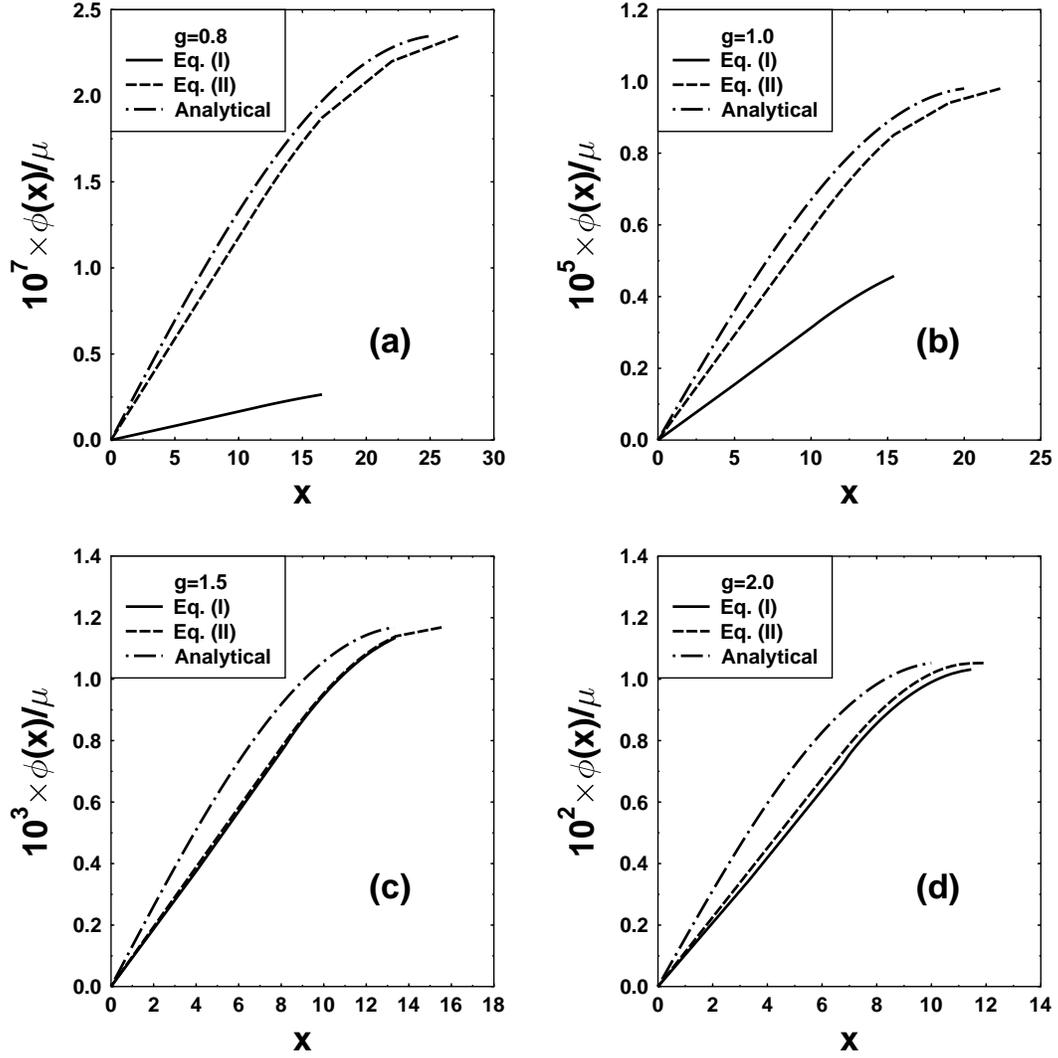,angle=0,width=1.0\linewidth}}}
\vspace{1truein}
\caption{The gap amplitude $\phi(x)$ as function of the variable $x$.
The solid line is a solution of Eq. (I) and the dashed line is a solution of   Eq. (II). 
The dot-dashed line is an analytic solution to the leading order of coupling constant $g$.}
\label{fig3}
\end{figure}
%%%%%%%%%%%%%%%%%%%%%%%%%%%%%%%%%%%%%%%%%%%%%%%%%%%%%%%%%%%%%%
%%%%%%%%%%%%%%%%%%%%%%%%%%%%%%%%%%%%%%%%%%%%%%%%%%%%%%%%%%%%%
% fig4
%
\begin{figure}
\centerline{\mbox{\epsfig{file=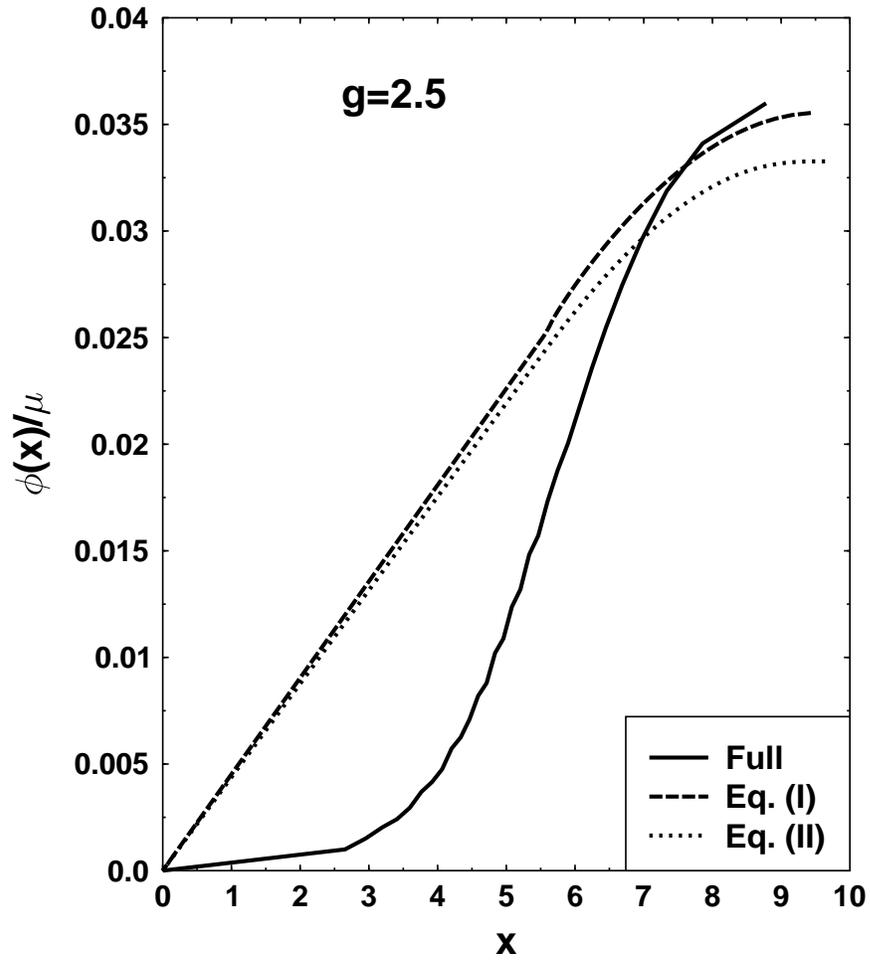,angle=0,width=0.8\linewidth}}}
\vspace{1truein}
\caption{The gap amplitude $\phi(x)$ as function of the variable $x$ with coupling constant $g=2.5$.
The solid line is a solution of the full gap equation.
The dashed line is a solution of Eq. (I) as well as
the doted line is a solution of Eq. (II).}
\label{fig4}
\end{figure}
%%%%%%%%%%%%%%%%%

% fig5
\begin{figure}
\centerline{\mbox{\epsfig{file=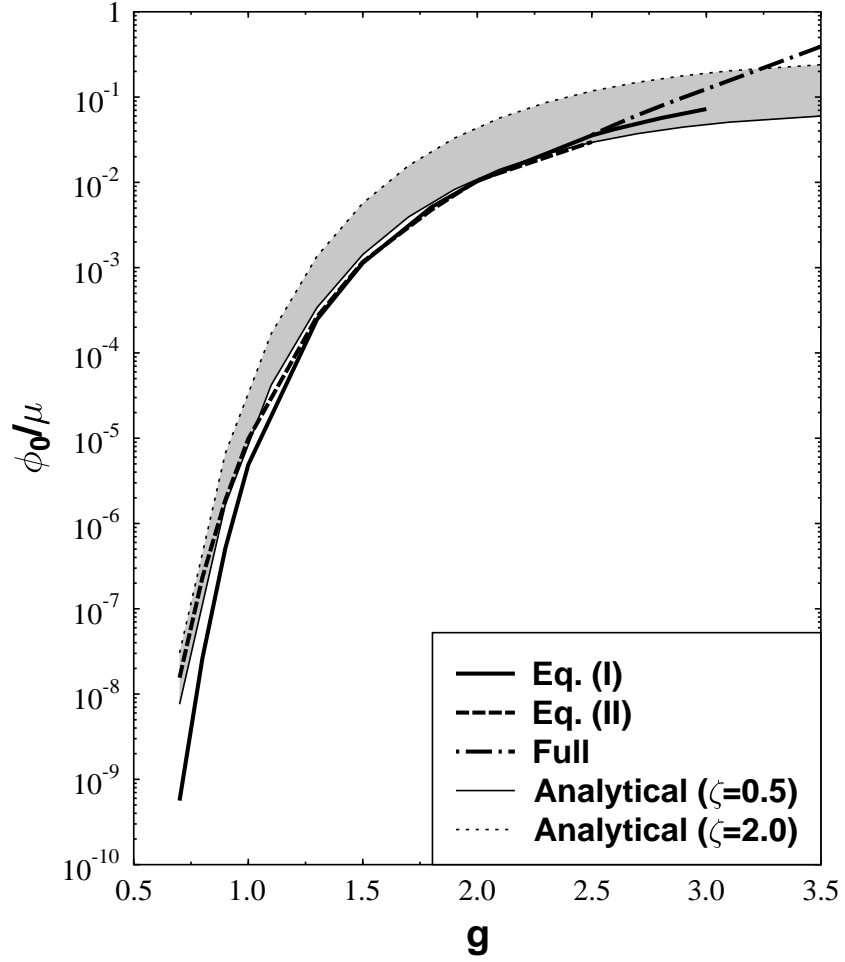,angle=0,width=0.8\linewidth}}}
\vspace{1truein}
\caption{The color superconductivity gap amplitude $\phi_0/\mu$ versus the coupling constant $g$ using several approximations. 
The solid line is a solution of Eq. (I) and the dashed line is a solution of 
Eq. (II) while the dotted-dashed line is a solution of the full gap equation. 
The light solid and dotted lines are the analytic solutions to leading order of coupling constant $\overline{g}$ multiplied by 
prefactor coefficient $\zeta=0.5$ and $\zeta=2.0$, respectively.}
\label{fig5}
\end{figure}
%%%%%%%%%%%%%%%

% fig6
\begin{figure}
\centerline{\mbox{\epsfig{file=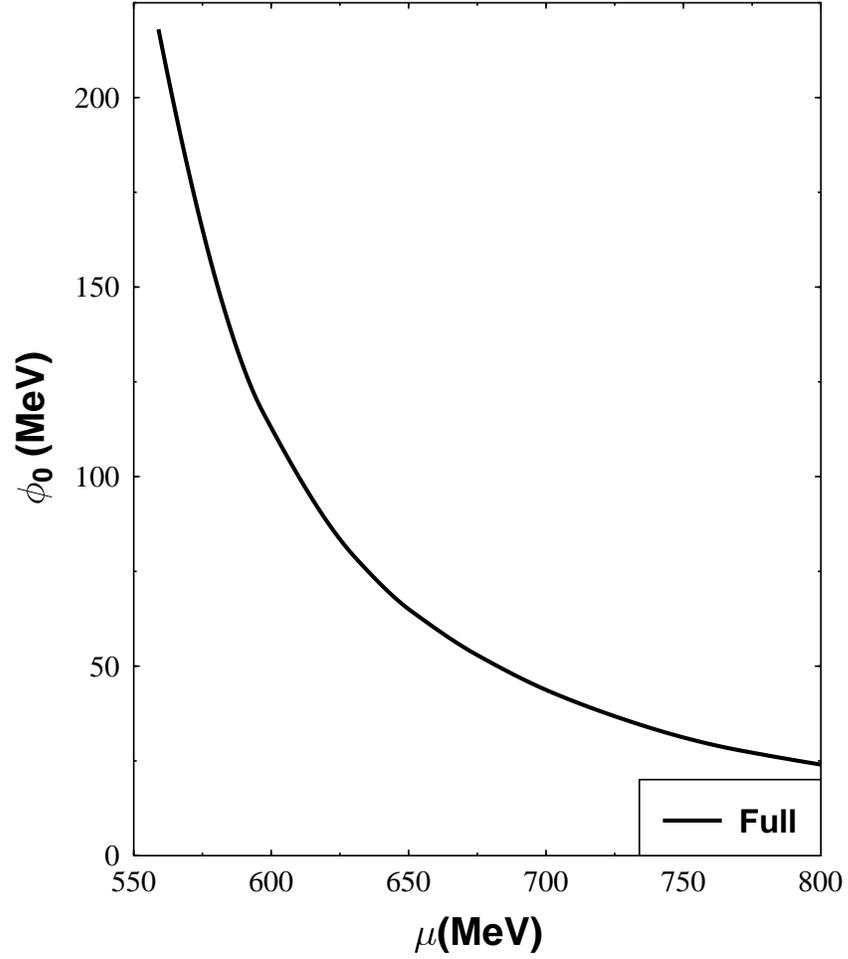,angle=0,width=0.8\linewidth}}}
\vspace{1truein}
\caption{The color superconductivity gap amplitude $\phi_0$ versus the chemical potential $\mu$ for the full gap equation. 
The coupling constant $g$ is identified with a standard logarithmic running coupling with QCD scale $\Lambda_{QCD}=400$ MeV.} 
\label{fig6}
\end{figure}

\end{document}